\documentclass[sigconf]{acmart}

\usepackage[most]{tcolorbox}
\usepackage{adjustbox}
\usepackage{float}
\usepackage{placeins}
\usepackage{array}

\usepackage{enumitem}
\usepackage{comment}

\usepackage{pgfplots}
\pgfplotsset{compat=1.18}
\usepackage{pgffor}

\AtBeginDocument{%
 }

\copyrightyear{2026}
\acmYear{2026}
\setcopyright{cc}
\setcctype{by}
\acmConference[CHASE'26]{19th International Conference International Workshop on Cooperative and Human Aspects of Software }{April 12--18, 2026}{Rio de Janeiro, Brazil}
\acmBooktitle{19th International Conference International Workshop on Cooperative and Human Aspects of Software (CHASE'26), April 12--18, 2026, Rio de Janeiro, Brazil}
\acmDOI{10.1145/3794860.3794911}
\acmISBN{979-8-4007-2484-8/2026/04}


\begin{document}
\title{Governance in Practice: How Open Source Projects Define and Document Roles}

\author{Pedro Oliveira}
\email{Pedro.Oliveira@nau.edu}
\affiliation{%
 \institution{Northern Arizona University}
 \city{Flagstaff}
 \state{AZ}
 \country{USA}
}

\author{Tayana Conte}
\email{tayana@icomp.ufam.edu.br}
\affiliation{
    \institution{Federal University of Amazonas}
    \city{Manaus}
    \state{Amazonas}
    \country{Brazil}
}

\author{Marco Gerosa}
\email{marco.gerosa@nau.edu}
\affiliation{%
 \institution{Northern Arizona University}
 \city{Flagstaff}
 \state{AZ}
 \country{USA}
}

\author{Igor Steinmacher}
\email{igor.steinmacher@nau.edu}
\affiliation{%
 \institution{Northern Arizona University}
 \city{Flagstaff}
 \state{AZ}
 \country{USA}
}

\renewcommand{\shortauthors}{Oliveira et al.}

\begin{abstract}
Open source software (OSS) sustainability depends not only on code contributions but also on governance structures that define who decides, who acts, and how responsibility is distributed. We lack systematic empirical evidence of how projects formally codify roles and authority in written artifacts. 
This paper investigates how OSS projects define and structure governance through their \texttt{GOVERNANCE.md} files and related documents. We analyze governance as an institutional infrastructure, a set of explicit rules that shape participation, decision rights, and community memory. We used Institutional Grammar to extract and formalize role definitions from repositories hosted on GitHub. We decompose each role into scope, privileges, obligations, and life-cycle rules to compare role structures across communities. 
Our results show that although OSS projects use a stable set of titles, identical titles carry different responsibilities, and different labels describe similar functions, which we call role drift. Still, we observed that a few actors sometimes accumulate technical, managerial, and community duties. 
By understanding authority and responsibilities in OSS, our findings inform researchers and practitioners on the importance of designing clearer roles, distributing work, and reducing leadership overload to support healthier and more sustainable communities.

\end{abstract}

 
\maketitle

\section{Introduction}
Open source software (OSS) plays a central role in the digital infrastructure that sustains modern computing. Over 90\% of commercial software includes OSS components~\cite{nagle2024value}, and nearly all major technology platforms depend on them~\cite{conti2024creators}. As the role of OSS expands, governance has emerged as a key factor related to project sustainability. Both industry and institutional sources emphasize that effective governance is central to sustaining healthy OSS communities ~\cite{redhat2022enterprise, ospo2023ggi}. Governance defines how decisions are made, who has authority, and how responsibilities are distributed. 

The governance structures influence contributor retention, diversity, and resilience~\cite{goggins2021making, trinkenreich2023belong, steinmacher2018let}. Several OSS projects still rely on informal or opaque governance practices, often centered on a few long-standing maintainers~\cite{guizani2021long, linaaker2024sustaining, raman2020stress}. When these governance mechanisms remain undefined, projects risk stagnation, contributor burnout, and a decline in trust. These issues have been recognized as systemic threats to the sustainability of the OSS ecosystem ~\cite{linuxfoundation2023annual}.

Despite its recognized importance, OSS governance remains empirically underexplored. Existing guidance often derives from gray literature (e.g., foundation templates, blog posts, and practitioner)~\cite{eghbal2016roads, neary2020governance, apacheGovernance, eclipseGovernance, todoGovernance, ospo2023ggi}, rather than from a systematic analysis of real-world governance artifacts. 


Empirical understanding of how governance is defined and varies across projects remains limited. Prior work focusing on individual projects, such as the case study of the \texttt{data.table} community~\cite{oliveira2025governance}, provides valuable qualitative insights into the consequences of restructuring projects through governance models, but lacks a cross-project perspective that captures the diversity of governance structures across the OSS landscape. 


As a result, we still lack a clear understanding of how authority and responsibility are documented in OSS projects. Without a systematic analysis of these governance documents, the authority distribution and how contributors gain or lose influence within different organizational structures remain unclear. 


In this paper, we characterize how OSS projects define, structure, and differentiate roles in their written governance models. Governance artifacts (such as \texttt{GOVERNANCE.md} files) offer an explicit record of how communities describe participation. We focus on these textual traces to understand how projects formalize collaboration, transforming tacit community norms into written structures that define how contributors coordinate their work.

To achieve this, we analyze a sample of GitHub repositories. From each repository, we identify governance files and extract all documented roles into a structured representation grounded in Institutional Grammar (IG)~\cite{crawford1995grammar, siddiki2011dissecting, sen2022cui}. We then map these roles to a catalog of technical, managerial, and interpersonal skills and combine computational analysis with manual interpretive clustering to compare how communities differentiate their roles. 


Our analyses reveal that OSS projects vary widely in how they label and define roles. The same title often carries distinct responsibilities across projects, while similar duties are often referred to under different names. These inconsistencies reveal a persistent misalignment of terminology, a lack of definitions, and overlapping responsibilities. By systematically analyzing these divergences, our study provides both an empirical foundation and practical guidance for maintainers, foundations, and researchers seeking to design models of participation in OSS governance, with the broader aim of fostering healthier OSS communities.



\section{Related Work}
OSS communities are living examples of large-scale collaboration. Since early studies, OSS has been conceptualized as a form of community of practice ~\cite{lave1991learning}, where learning, legitimacy, and authority emerge through participation, rather than formal appointment. Early governance models, such as the benevolent dictator paradigm or consensus-based meritocracies, depended on informal social contracts, technical reputation, and sustained contribution instead of codified policy~\cite{boehm2019emergence, ritvo2017challenges, linaaker2019community}. 

As OSS projects grow in scale and complexity, empirical studies indicate that informal governance arrangements often become insufficient for ensuring accountability, transparency, and continuity~\cite{jensen2010governance, yin2022open}. This leads to the institutionalization of OSS governance, where communities formalize norms and practices into explicit structures like charters, and artifacts to manage coordination challenges and sustain development~\cite{chakraborti2024we, demil2006neither, eghbal2016roads}.

\subsection{Roles, Pathways, and Belonging in OSS}

Understanding governance requires understanding the social trajectories that sustain open source communities. Contributors rarely assume authority immediately. They progress from peripheral participation to more central forms of involvement through a combination of learning, recognition, and sustained interaction~\cite{trinkenreich2020hidden, song2021role, zhou2014will}. These trajectories are shaped not only by technical expertise, but also by social and communicative engagement, as contributors' progress depends on their ability to build relationships, exchange feedback, and gain recognition within the community~\cite{trinkenreich2020hidden, steinmacher2014attracting, barcomb2018uncovering, zhou2014will}. 

More recent work reframes these challenges as affective rather than procedural. Contributors' sense of belonging and identification with the community was evidenced as a stronger predictor of retention than technical proficiency alone~\cite{trinkenreich2023belong}. Collectively, these studies highlight that governance defines the visible and invisible pathways through which contributors participate inside the project.


\subsection{The (Lack of) Documented Governance}

Despite the extensive body of research describing OSS roles, pathways, and collaboration practices, few studies have examined governance as it is documented in project repositories. Existing work has mostly characterized how contributors gain authority or how leadership structures evolve, but has rarely investigated the governance artifacts that define these arrangements~\cite{o2007emergence, jensen2010governance}. Recent work has begun to explore the consequences of formalizing governance models, Oliveira et al.~\cite{oliveira2025governance} showed how rewriting a project's governance file reshaped decision-making patterns and community health in the \texttt{data.table} ecosystem.

Although multiple studies discuss roles such as maintainers, committers, or core contributors~\cite{barcomb2018uncovering, trinkenreich2020hidden}, little is known about how such roles (e.g., associated privileges, responsibilities, and decision rights) are formally defined in writing. We still lack an empirical account of what responsibilities and decision structures OSS projects codify, and how these vary across ecosystems.

This study addresses that gap by analyzing governance documents across a diverse sample of OSS projects. Through a large-scale, cross-ecosystem comparison, we reveal how authority, responsibility, and participation are codified in writing, bridging sociotechnical theory, which views governance as both a social process and a technical artifact, with textual evidence to reveal how communities formalize governance in practice.



\subsection{Institutional Grammar}
Institutional Grammar (IG) provides a structured framework for analyzing how governance rules are expressed in written form. IG conceptualizes institutions as systems of statements that specify who may or must do \textit{what, under which conditions, and with what consequences}~\cite{crawford1995grammar}. Rather than treating governance as an abstract structure alone, IG focuses on the language through which authority, permissions, and responsibilities are formally articulated.

To enable systematic analysis, each institutional statement can be decomposed into syntactic components, commonly summarized as Attribute (actor), Deontic (permission, obligation, or prohibition), Aim/Object (action), Conditions (context), and Or-else (sanction). This decomposition makes heterogeneous governance texts comparable by transforming qualitative rules into structured, analyzable elements ~\cite{siddiki2011dissecting}.

In this study, we adopt a lightweight, role-centric interpretation of IG tailored to software projects. Instead of decomposing every sentence at the clause level, we operationalize the grammar into four practical dimensions: scope, privileges, obligations, and promotion/demotion criteria. These dimensions correspond to the Attribute-Deontic-Aim-Condition structure while remaining tractable for large-scale comparative analysis. This adaptation enables consistent extraction and comparison of how projects formalize authority in governance documents.

\section{Research Method}

We conducted a qualitative analysis of how OSS projects define, structure, and differentiate roles through written governance artifacts. Governance files represent one of the publicly accessible traces of how communities formalize who participates, the responsibilities they have, and under what conditions. We systematically collected governance files from multiple OSS projects on GitHub, coded their contents to identify explicit role definitions, and compared how authority, responsibility, and participation are described across projects.

Following the qualitative approaches discussed by Easterbrook et al.~\cite{easterbrook2008selecting}, our inductive analysis centers on three interconnected lenses: how governance artifacts \textbf{codify roles} and delineate participation; how they \textbf{distribute responsibilities and decision rights} across these roles; and how they \textbf{differentiate} or blur boundaries between technical, managerial, and symbolic functions.

The research process followed an iterative approach combining exploration, structured extraction, and interpretive analysis. We began by familiarizing ourselves with governance files to identify recurring structures and language. Guided by these observations, we applied systematic data extraction, skill mapping, and successive clustering, refining each step through discussion and validation. 

\subsection{Researcher Positionality and Reflectivity}
This study was conducted by researchers with long-standing involvement in OSS communities and experience studying socio-technical collaboration. All the authors have experience in conducting empirical studies on OSS community governance, onboarding, and sustainability. This insider familiarity offered an interesting understanding of governance language and practices, but also carried the risk of interpretive bias.

To mitigate the potential bias, we adopted a reflexive stance throughout the analysis. Coding and interpretation were discussed collaboratively, with deliberate attention to negative cases and staying close to the textual evidence rather than presumed norms of "good governance." Iterative peer debriefing among the authors helped ensure that decisions about role categorization and skill mapping reflected the actual data. Our goal was not to judge whether projects governed "well", but to understand how authority and responsibility were codified in practice.

\subsection{Finding Candidate Projects}
Our population consists of OSS projects hosted on GitHub, which represents one of the largest ecosystems for software development and collaboration~\cite{github_about, github_100million}. To identify suitable candidates for our analysis, we focused on repositories that (i) are publicly available, (ii) explicitly define an open-source license, and (iii) have community recognition as reflected by stargazer count. 

Based on these criteria, we collected a license-stratified set of projects on GitHub. Our selection is grounded in the recent license popularity rankings reported by GitHub’s Innovation Graph~\cite{licenseRanking}. We included the following top licenses in our sampling: MIT, GPL-2.0, GPL-3.0, Apache-2.0, BSD-2-Clause, BSD-3-Clause, MPL-2.0, LGPL, EPL-2.0, CC0-1.0, and AGPL-3.0. We then retrieved 1,000 repositories licensed under each category, ordered them by stargazer count in descending order, as showed in Table ~\ref{tab:projectsWithGovernance}.

After the project retrieval, we checked the presence of governance files. Following GitHub's official documentation~\cite{github_governance_doc}, which recognizes \texttt{GOVERNANCE.md} as a file that describes project governance, we restricted our search to files whose names contained the term "governance". The search was executed recursively, at any folder depth within the repository tree, in any file format. We dismissed the projects in which we did not detect a governance file.

\begin{table}[!htbp]
      \centering
      \caption{Projects with Governance Files.}
      \small
      \begin{tabular}{
        l 
        l 
        r@{\,}l 
        r@{\,}l 
      }
       \hline \textbf{License} & \textbf{Projects} & \multicolumn{2}{l}{\textbf{Files}} & \multicolumn{2}{l}{\textbf{Governance}} \\
        \textbf{Type} & \textbf{Collected} & \multicolumn{2}{l}{\textbf{Detected}} & \multicolumn{2}{l}{\textbf{Files w/ Content}} \\\hline
        Apache-2.0 & 1,000 & 40 & (4.0\%) & 30 & (3.0\%) \\
        MIT & 1,000 & 10 & (1.0\%) & 7 & (0.7\%) \\
        AGPL-3.0 & 1,000 & 7 & (0.7\%) & 5 & (0.5\%) \\
        BSD-3-Clause & 1,000 & 6 & (0.6\%) & 3 & (0.3\%) \\
        BSD-2-Clause & 1,000 & 3 & (0.3\%) & 3 & (0.3\%) \\
        MPL-2.0 & 1,000 & 3 & (0.3\%) & 3 & (0.3\%) \\
        LGPL-2.1 & 1,000 & 2 & (0.2\%) & 1 & (0.1\%) \\
        CC0-1.0 & 1,000 & 1 & (0.1\%) & 1 & (0.1\%) \\\hline
        \textbf{Total} & \textbf{8,000} & \textbf{72} & \textbf{(0.90\%)} & \textbf{54} & \textbf{(0.67\%)} \\\hline
      \end{tabular}
  \label{tab:projectsWithGovernance}
\end{table}

\subsection{Extracting Governance Topics}

Once candidate projects were identified, we analyzed the themes in the governance files using a two-stage qualitative coding process following Saldana's guidelines for iterative coding~\cite{saldana2021coding}. In the first cycle, we conducted an open reading of each governance file and recorded all distinct topics explicitly mentioned in the text. Examples included references to merging rights, release authority, review obligations, quorum requirements, and funding arrangements. Each element was coded at the level of granularity in which it appeared, preserving the original terminology and scope.

In the second cycle, we compared and clustered related subtopics into broader categories that reflected recurring governance mechanisms. For instance, provisions about merging rights, release procedures, and security disclosures were grouped under a category representing core technical processes. Finally, we consolidated these categories into a smaller set of topics that summarize the main dimensions addressed by governance files across projects. This iterative process yielded six high-level topics that structure our subsequent analysis:

\begin{itemize}[leftmargin=*]
  \item \textbf{Organizational Structure} - how roles are defined, responsibilities distributed, and authority transferred.
  \item \textbf{Decision-Making} - the mechanisms for collective choices, including voting and elections.
  \item \textbf{Core Processes} - technical workflows such as release management, contribution pipelines, and security disclosures.
  \item \textbf{Community and Communication} - norms for interaction, including codes of conduct and official channels.
  \item \textbf{Project Context} - institutional affiliations, licensing, and maturity signals.
  \item \textbf{Compensation Schemes} - financial arrangements, funding sources, and disbursement rules.
\end{itemize}

By deriving topics through this bottom-up grouping, we ensured that our coding was grounded in the content of governance files rather than in a predefined structure. This method allowed us to compare heterogeneous documents while being sensitive to project-specific terminology. The frequencies for each topic are reported in Table~\ref{tab:topicFrequency}. It is important to note that frequencies were calculated at the document level: a topic was considered present in a file if at least one of its associated subtopics appeared in the text.

\begin{table}[ht]
  \centering
  \caption{Topics Approached by Governance Documents.}
  \small
  \begin{tabular}{p{5cm} @{\hspace{0.8cm}} p{0.8cm}<{\raggedleft} p{1cm}<{\raggedright}}
    \hline\textbf{Section} & \multicolumn{2}{c}{\textbf{Frequency}} \\\hline
    Organizational Structure & 54 & (100\%) \\
    Decision Making & 53 & (98\%)  \\
    Core Processes & 49 & (90\%)  \\
    Community and Communication & 43 & (80\%)  \\
    Project Context & 32 & (59\%)  \\ 
    Compensation Schemes & 7  & (13\%)  \\\hline
  \end{tabular}
  \label{tab:topicFrequency}
\end{table}

\subsection{Extracting Role Information}
Among the six topics, organizational structure emerged as the one documented in all projects and the most directly related to our research questions. Governance files varied in length and emphasis, but all specify, implicitly or explicitly, a set of roles and the distribution of authority among them. We conducted a detailed extraction focused specifically on this dimension.


In this step, we aimed to move from unstructured textual descriptions to structured and comparable records of governance roles. To achieve this, we designed an extraction template grounded in Institutional Grammar~\cite{crawford1995grammar, siddiki2011dissecting, sen2022cui}, a framework for decomposing institutional statements into distinct analytical components. IG models governance as a composition of Attribute (the actor or role), Deontic (the normative modality indicating permissions or obligations), Aim/Object (the activity or domain of action), Conditions (the circumstances or qualifications under which a rule applies), and Or-else (sanctions). 

Following this model, we coded each identified role according to four aspects that operationalize these elements in the context of OSS governance.

\begin{itemize}[leftmargin=*]
  \item \textbf{Scope of Responsibility} - The activities, domains, or areas the role is accountable for (e.g., "responsible for releases", "maintains project infrastructure").
  \item \textbf{Privileges} - The explicit powers granted to the role (e.g., merge or commit rights, voting eligibility, authority to approve pull requests, or to represent the project externally).
  \item \textbf{Responsibilities} - The duties or expectations associated with the role (e.g., reviewing contributions, mentoring newcomers, participating in decision-making processes).
  \item \textbf{Promotion and Demotion Criteria} - The rules that determine how individuals acquire or lose a role (e.g., nomination and election, activity thresholds, appointment by a committee, or voluntary resignation).
\end{itemize}

Each of these aspects captures a different dimension of authority: scope defines what is governed, privileges indicate the levers of power, obligations reveal the expectations attached to authority, and entry/exit criteria regulate access to these positions. Together, they enable us to reconstruct governance as a set of titles and as a structured system of rules for participation and control.

We annotated governance files using this template and coded role definitions only when explicit textual evidence was present. If the role's privilege, obligation, or promotion/demotion criteria were not stated, the corresponding field was marked as "absent" rather than inferred. The output of this step, available in the replication package \cite{paper_replication}, was a role-centric dataset, where each entry represents a single role instance described in a governance document.

\subsection{Mapping Titles to Talents}
As our analysis of governance documents progressed, we noted that the role names drifted across projects. The same label could signal very different responsibilities, while different labels could denote nearly identical positions. Terms such as "Owner", "Project Lead", or "Core Developers" were occasionally used to describe almost the same responsibilities.
This drift made it problematic to rely on titles alone for comparison. Treating the title as equivalent risked conflating distinct responsibilities, while treating them as different risked ignoring similarities. To resolve this, we decided to move beyond focus on the skills encoded in governance provisions. 


For this step, we adapted the open-source skills catalog proposed by Liang et al.~\cite{liang2022understanding} to fit our analytical goals. Table~\ref{tab:skillsDefinition} presents our adapted version, with the respective IDs for each skill. Each extracted role from the governance files was mapped to this catalog using the evidence we collected, specifically, the scope, privileges, obligations, and promotion/demotion criteria.

\begin{table}[!htbp]
  \centering
  \caption{Skills Catalog Description~\cite{liang2022understanding}.}
  \label{tab:skillsDefinition}
  \footnotesize
  \begin{tabular}{|p{0.6cm}p{2.8cm}|p{0.6cm}p{2.8cm}|}
    \hline
    \textbf{ID} & \textbf{Name} & \textbf{ID} & \textbf{Name} \\
    \hline

    \multicolumn{4}{c}{\textbf{Technical Skills (TS)}} \\\hline
    T1 & Programming & T4 & DevOps \\
    T2 & Software Engineering & T5 & Domain \\
    T3 & Technologies & T6 & Version Control Systems \\\hline

    \multicolumn{4}{c}{\textbf{Working Styles (W)}} \\\hline
    W1 & Excellence & W3 & Organized \\
    W2 & Available &  &  \\\hline

    \multicolumn{4}{c}{\textbf{Problem Solving (P)}} \\\hline
    P1 & Creative & P3 & Analytical \\
    P2 & Initiative &  &  \\\hline

    \multicolumn{4}{c}{\textbf{Contribution Type Skills (CT)}} \\\hline
    CT1 & Bug triaging & CT4 & Documentation \\
    CT2 & Bug reporting & CT5 & Visual Design \\
    CT3 & Code Review & CT6 & Translation \\\hline

    \multicolumn{4}{c}{\textbf{Project-Specific Skills (PSS)}} \\\hline
    PSS1 & Purpose & PSS3 & Processes \\
    PSS2 & Organization &  &  \\\hline

    \multicolumn{4}{c}{\textbf{Interpersonal Skills (IS)}} \\\hline
    IS1 & Kind & IS4 & Giving help \\
    IS2 & Communication & IS5 & Conflict Resolution \\
    IS3 & Asking for help & IS6 & Collaboration \\\hline

    \multicolumn{4}{c}{\textbf{External Relations (ER)}} \\\hline
    ER1 & Stakeholders & ER3 & Licenses \\
    ER2 & Marketing & ER4 & Funding \\\hline

    \multicolumn{4}{c}{\textbf{Management (M)}} \\\hline
    M1 & Community & M4 & Delegating \\
    M2 & Project & M5 & Time \\
    M3 & Planning &  &  \\\hline

    \multicolumn{4}{c}{\textbf{Characteristics (C)}} \\\hline
    C1 & Adventurous & C6 & Reliable \\
    C2 & Open-minded & C7 & Persevering \\
    C3 & Patient & C8 & Diligent \\
    C4 & Adaptable & C9 & Self-aware \\
    C5 & Curious &  &  \\\hline

  \end{tabular}
\end{table}

This reframing allowed us to analyze the distance between roles that shared the same name. For example, two distinct "Maintainer" roles could now be compared in terms of their skill profiles. If their bundles overlapped only partially, the distance between them was large; if they converged on similar competencies, the distance was small. To formalize this comparison, we applied a series of unsupervised clustering experiments using the 45-dimensional binary skill matrix extracted from governance files. Each role instance was represented as a vector of skill presences (Table~\ref{tab:skillsDefinition}), standardized with scikit-learn's StandardScaler function~\cite{scikit_learn_standardscaler}. 

\begin{figure}[!htbp]
  \centering
  \includegraphics[width=1\linewidth]{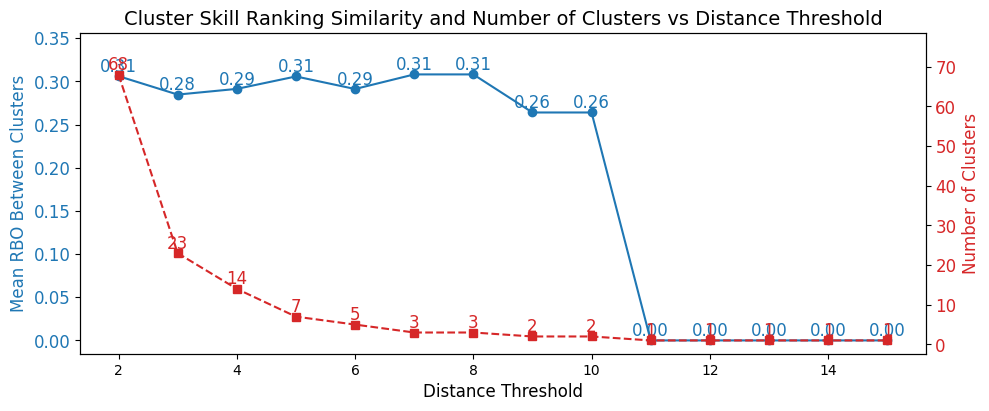}
  \caption{Role Automatic Clustering.}
  \label{fig:clusteringMetricsOutput}
  \Description{A dual-axis line chart shows how cluster similarity and number of clusters change as the distance threshold increases from 2 to 15. The blue solid line (left axis) represents mean Rank-Biased Overlap (RBO) between clusters. It remains relatively stable between 0.26 and 0.31 from thresholds 2 to 10, then drops to 0.00 at threshold 11 and remains at zero through 15. The red dashed line (right axis) represents the number of clusters. It decreases sharply from 68 clusters at threshold 2 to 23 (3), 14 (4), 7 (5), 5 (6), 3 (7–8), 2 (9–10), and 1 cluster from threshold 11 onward. The figure shows that increasing the threshold progressively merges clusters until all roles collapse into a single cluster, after which no intra-cluster similarity is measurable.}
\end{figure}

We tested multiple algorithms (K-Means, Agglomerative Clustering, and DBSCAN) to identify potential groupings. Role-to-role similarity was computed using Euclidean distance, and cluster coherence was evaluated with Rank-Biased Overlap (RBO) between ranked skill frequency profiles at different thresholds (Figure~\ref{fig:clusteringMetricsOutput}). However, the results showed high sensitivity to parameter choices and low stability across thresholds. At smaller distance cutoffs, agglomerative clustering produced many micro-clusters, while higher thresholds merged roles into broad, uninformative categories. 

A closer investigation revealed that the reason was that several governance roles were not atomic but composite, aggregating multiple smaller responsibilities that, in other projects, were distributed across distinct roles. For example, one project's "Maintainer" could simultaneously act as reviewer, release manager, community leader, and triager. 

These hybrid roles inflated their skill bundles, causing them to appear artificially similar to multiple categories at once and undermining the convergence of the clustering. Recognizing this limitation, we decided not to overinterpret the unstable clusters and instead advance toward another analytical approach capable of capturing these composite role structures.

\subsection{Manual Role Clustering Interpretation}
The next step was a manual interpretive clustering process, recognizing that governance roles are complex social artifacts. They are embedded in community norms, historical trajectories, and project-specific practices that resist purely computational classification. Our interpretive clustering unfolded in three steps.
\begin{enumerate}[leftmargin=*]
  \item We revisited the role definitions extracted from governance files together with their mapped skills bundles. Rather than treating skills as independent features to be automatically grouped, we examined how they were expressed in context: what combination of privileges, obligations, and responsibilities they represented within each project.
  
  \item We compared roles across projects, looking for overlaps and connections. When two roles showed similarities, we grouped them into the same cluster. We prioritized classification into the most prominent or dominant cluster, then recorded cross-links to secondary or tertiary clusters where overlaps existed. This enabled us to acknowledge ambiguity without fragmenting the dataset into an explosion of micro-categories. For example, a "Maintainer" who also performed triage and release tasks was primarily assigned to the Core Maintainer cluster, but linked to Triage and Release Manager as secondary roles
  
  \item We consolidated the assignments into a set of recognizable clusters. These included Contributor, Core Maintainer, Steering/Leadership, Reviewer, Triage, User, and Project-Specific Roles, along with less frequent clusters such as Advocacy or Emeritus positions. The frequency of these categories across projects is reported in Table~\ref{tab:manualRoles}, which reveals both dominant role types (e.g., Contributor and Maintainer) and symbolic or transitional ones (e.g., Emeritus Maintainer).
\end{enumerate}

\begin{table}[!htbp]
  \centering
  \caption{Manual Clustering Result.}
  \label{tab:manualRoles}
  \small
  \begin{tabular}{p{2.3cm}r @{\hspace{0.6cm}} p{2.5cm}r}
    \hline
    \textbf{CLUSTER} & \textbf{SIZE} & \textbf{CLUSTER} & \textbf{SIZE} \\
    \hline

    Core Maintainer & 50 & Advocacy & 4 \\
    Contributor & 29 & Emeritus Maintainer & 4 \\
    Steering & 20 & Triage & 3 \\
    User & 12 & Reviewer & 2 \\
    Project Specific & 9 & Committer & 1 \\
    Owner & 8 &  &  \\

    \hline
    \multicolumn{3}{r}{\textbf{TOTAL}} & \textbf{133} \\
    \hline
  \end{tabular}
\end{table}


The interpretive clustering was conducted in one intensive session involving three of the authors. Two of them have over a decade of experience studying OSS sustainability, contributing extensive domain knowledge to the process. Each role identified per file was discussed collaboratively, one by one, examining its textual definition, skill composition, and possible alignment within the emerging clusters. Decisions were made only through unanimous agreement among all three researchers, ensuring that each categorization was grounded in shared understanding and justified reasoning.

To support our reasoning, we used a shared whiteboard to visualize the relationships among roles and their functional domains. This visualization served as a living artifact. As discussions unfolded, we mapped hierarchical dependencies and overlapping responsibilities. The evolving diagram enabled us to trace structural connections between roles and reveal latent governance patterns (e.g., similar titles with different responsibilities across projects).

\begin{figure*}[ht!]
  \centering
  \includegraphics[width=1\linewidth]{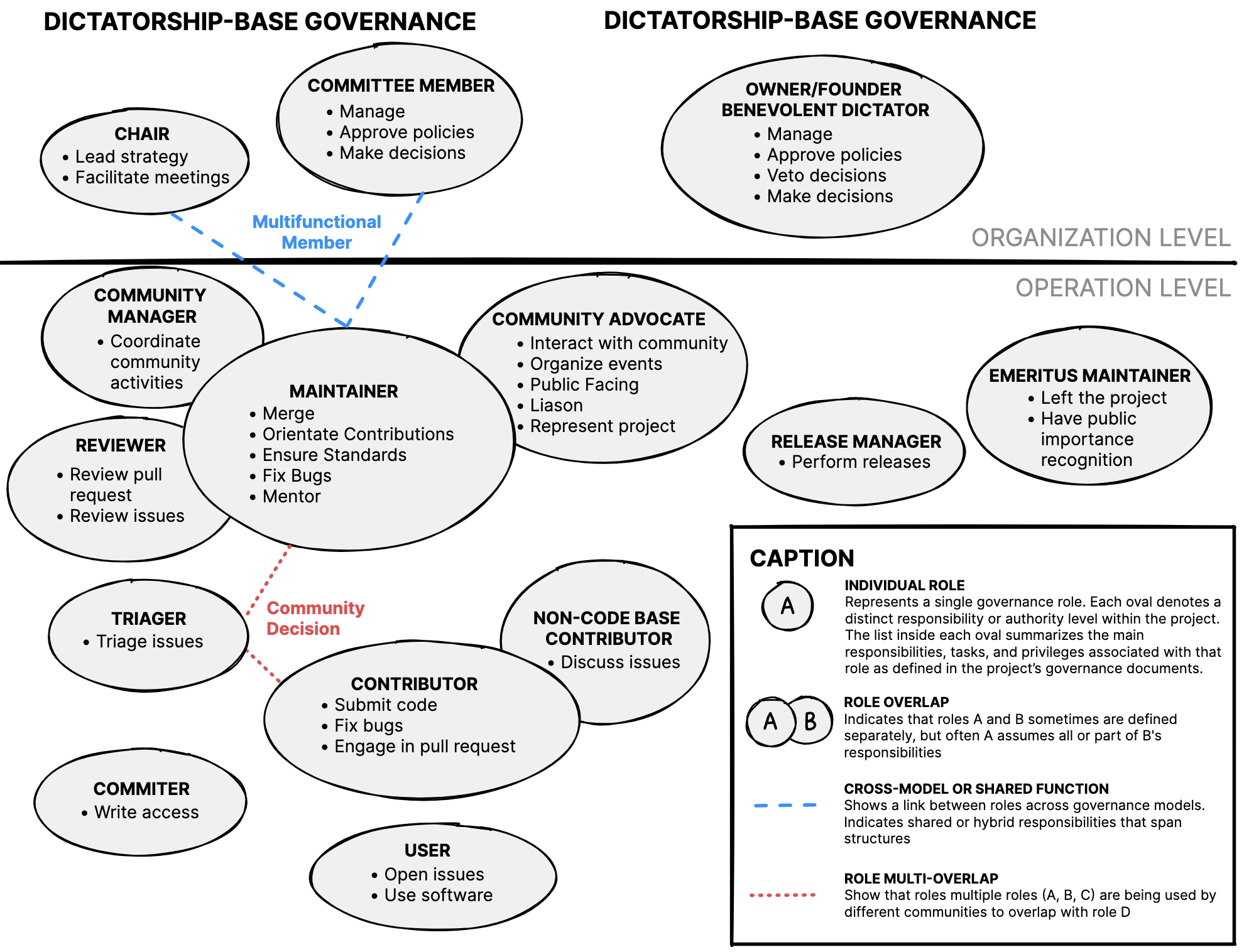}
  \caption{Roles Organization and Skill Distribution.}  \label{fig:rolesDefinition}
  \Description{The figure is a conceptual diagram illustrating governance roles and their relationships in open source projects. Roles are shown as labeled ovals containing short bullet lists of responsibilities. A horizontal line divides the diagram into two layers: Organization Level (top) and Operation Level (bottom). At the organization level, roles include Chair (leads strategy, facilitates meetings), Committee Member (manages, approves policies, makes decisions), and Owner/Founder/Benevolent Dictator (manages, approves policies, vetoes and makes decisions). At the operation level, the Maintainer is centrally positioned and includes responsibilities such as merging contributions, ensuring standards, fixing bugs, and mentoring. Surrounding operational roles include Reviewer (reviews pull requests and issues), Triager (triages issues), Committer (write access), Contributor (submits code, fixes bugs, engages in pull requests), Non-Code Base Contributor (discusses issues), User (opens issues and uses software), Release Manager (performs releases), Community Advocate (interacts with community, organizes events, public-facing), Community Manager (coordinates community activities), and Emeritus Maintainer (former maintainer with honorary recognition). Dashed and dotted lines indicate overlaps or multifunctional responsibilities across roles. The Maintainer connects to multiple roles, highlighting hybrid responsibilities. A legend explains visual conventions: individual roles (ovals), role overlap (shared responsibilities), cross-model or shared functions (blue dashed lines), and multi-role overlap (red dotted lines).

The figure conveys a layered governance structure in which strategic authority appears at the organizational level, while technical and community activities are concentrated at the operational level, with maintainers often bridging both layers.}
\end{figure*}

This interpretive approach was more effective, although more time-consuming, than automatic clustering because it could accommodate the composite nature of governance roles. By privileging a primary cluster while preserving cross-links, we balanced comparability with contextual sensitivity. Figure~\ref{fig:rolesDefinition} illustrates the resulting organization: roles are arranged into organizational and operational layers, with skill distributions showing both recurrent patterns and overlaps. All role definitions and cluster mappings are provided in our replication package~\cite{paper_replication} to ensure full traceability between source data and outcomes.

\section{Results}
We begin by presenting how OSS projects distribute responsibilities and authority across governance roles (Figure~\ref{fig:rolesDefinition}). Then we detail each governance dimension, unpacking how responsibilities, privileges, and skills are articulated in practice.

\subsection{Overlap and Differentiation of Responsibilities}
Figure~\ref{fig:rolesDefinition} provides a visual map of how governance documents describe the distribution of responsibilities across projects. Each bubble corresponds to a category derived from our interpretive analysis of governance files. The items inside the bubble summarize the main responsibilities explicitly associated with that role. The overlap/dashed lines between bubbles capture observed overlaps or dependencies across roles, respectively.



To ensure consistency, we classified roles into organizational or operational layers based on their formal decision authority rather than the technicality of their tasks. Roles were labeled organizational when governance documents granted them strategic or institutional powers (e.g., defining policies, voting on decisions, or setting project direction). Roles were labeled operational when they primarily executed, supported, or coordinated work without formal authority over project direction.

What emerges from this visualization is the diversity in how projects separate/combine responsibilities. In some projects, bubbles are distinct and specialized. Reviewers handle only code assessment, triagers manage bug queues, and release managers focus on distributing versions. Other projects, however, adopt multifunctional roles where bubbles merge.

The maintainer bubble, centrally positioned, often aggregates responsibilities that, in other projects, are distributed across several roles. A maintainer might simultaneously merge contributions, enforce coding standards, triage issues, mentor newcomers, and even act as a community representative. 

\subsection{What's in a Role? Skills Tell the Story}

\begin{figure}[ht!]
  \makebox[0pt][c]{%
    \begin{minipage}{1.05\columnwidth}
      \centering
      \includegraphics[width=\linewidth]{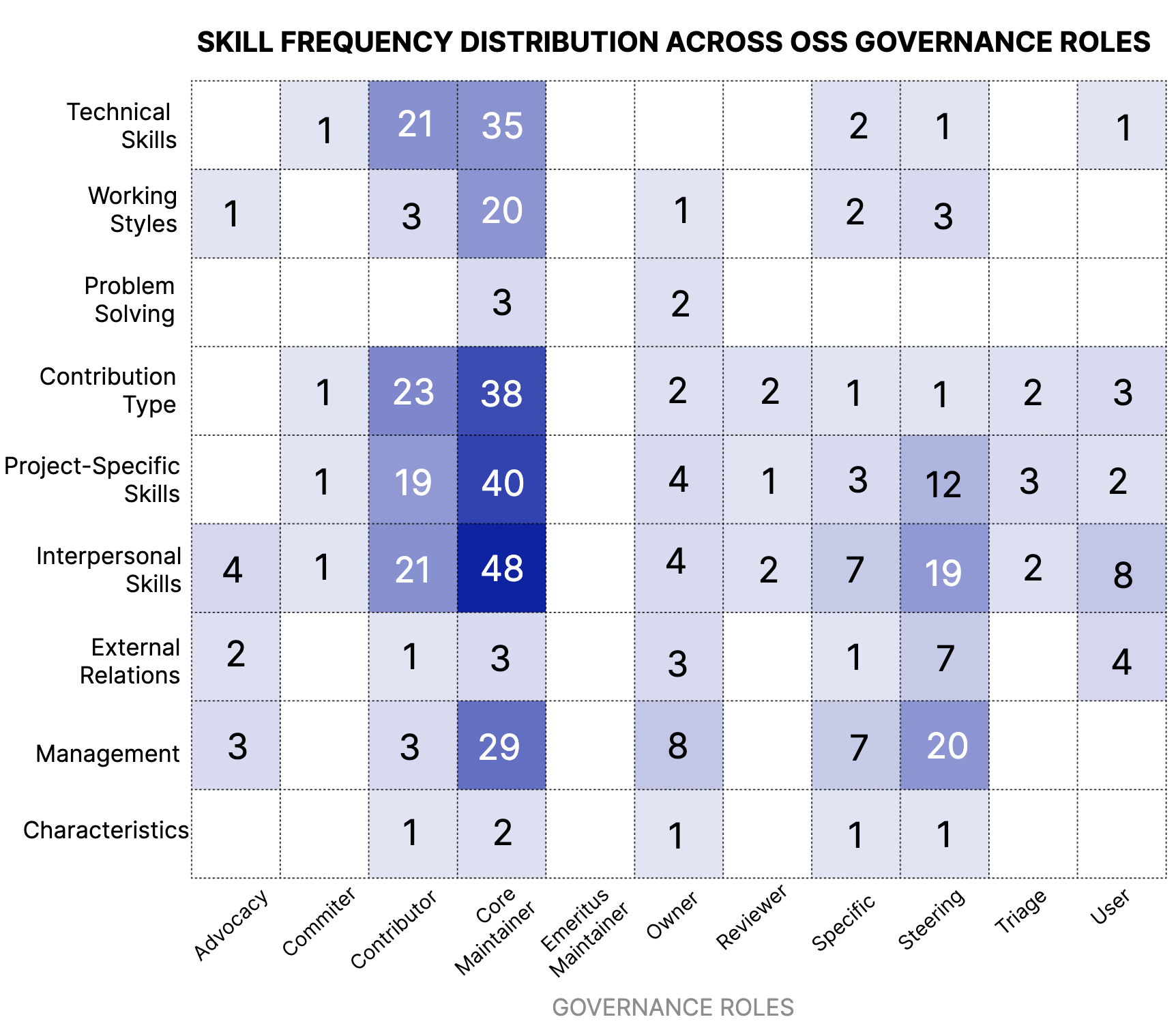}
      \caption{Skill Frequency Distribution Across Governance Roles.}
      \label{fig:skillDistribution}
    \end{minipage}
  }
  \Description{The figure is a heat map titled “Skill Frequency Distribution Across OSS Governance Roles.” The horizontal axis lists governance roles: Advocacy, Committer, Contributor, Core Maintainer, Emeritus Maintainer, Owner, Reviewer, Specific, Steering, Triage, and User. The vertical axis lists skill categories: Technical Skills, Working Styles, Problem Solving, Contribution Type, Project-Specific Skills, Interpersonal Skills, External Relations, Management, and Characteristics.

Each cell contains a numeric count representing how often a skill category appears in the definition of a given role. Darker blue shading indicates higher frequency.

Core Maintainers show the highest concentrations across multiple categories, especially Interpersonal Skills (48), Project-Specific Skills (40), Contribution Type (38), Technical Skills (35), and Management (29). Contributors also show high frequencies, including Contribution Type (23), Technical Skills (21), Interpersonal Skills (21), and Project-Specific Skills (19). Steering roles show elevated Management (20) and Interpersonal Skills (19). Owners emphasize Management (8) and Project-Specific Skills (4). Reviewers and Triagers show lower and more specialized counts, mainly in Contribution Type and Interpersonal Skills. Advocacy and User roles show comparatively low frequencies across categories.

The heat map illustrates that Core Maintainers and Contributors concentrate the broadest and most frequent skill distributions, while other roles display narrower or more specialized skill profiles.}
\end{figure}

Figure~\ref{fig:skillDistribution} reveals the distribution of skill categories across governance positions. The heat map shows how often each skill category appears in the definition of different governance roles. Darker shades indicate higher frequencies. The vertical axis lists the skill categories, while the horizontal axis represents governance role categories. Each cell reflects the number of occurrences of a skill type for a given role, illustrating which skills are most emphasized across the governance documents.

\subsubsection{Contributors: \textit{The Technical Backbone}}

The contributor role is the broadest and most balanced across technical and interpersonal skills. As shown in the graphs, contributors are commonly associated with programming (T1), bug reporting (CT2), code review (CT3), and documentation (CT4). This mix underscores that contributors may produce artifacts or engage socially through communication (IS2) and collaboration (IS6), often involving one or more of these activities rather than all of them.

For example, the \texttt{vitessio/vitess} governance file~\cite{vitess} captures this duality: contributors write code and documentation, are tasked with supporting new users and spreading the project through talks or advocacy. In other words, becoming a contributor is the entry point for participation, carrying productive or social obligations that may be fulfilled through one or more activities stated in the governance document. 

\begin{tcolorbox}[breakable, title={Org: vitessio | Repo: vitess}]
\textbf{Contributor}: 
Contributors are community members who contribute in concrete ways to the project. [...]
In addition to their actions as users, contributors may also find themselves doing one or more of the following: supporting new users (existing users are often the best people to support new users); reporting bugs; identifying requirements; providing graphics and web design; [...]
\end{tcolorbox}

Similar formulations appear in \texttt{bpftrace/bpftrace}~\cite{bpftrace}, \texttt{apolloconfig/apollo}~\cite{apollo}, and \texttt{Rdatatable/data.table}~\cite{datatable}, where contributors are described as open, entry-level participants engaging through both technical and community-facing activities.

\subsubsection{Core Maintainers: \textit{The Hybrid Role}}
No role is more central, or more composite, than core maintainer. Maintainers' profiles span across programming (T1), software engineering (T2), version control systems (T6), and documentation (CT4), while also emphasizing on management skills (M1, M2, M3) and interpersonal abilities like communication (IS2) and collaboration (IS6).

This breadth reflects their multi-dimensional mandate, ensuring technical quality and leading the community. Maintainers are not just coders. They are guardians of standards, planners of direction and vision, leaders, and often mentors. In the \texttt{marimo-team/marimo} project~\cite{marimo}, maintainers are (as they describe) the \textit{"ultimate authority"} over project direction, but their day-to-day may include triaging issues, reviewing pull requests, and maintaining CI pipelines, functions that in other projects are distributed across multiple roles.

\begin{tcolorbox}[breakable, title={Org: marimo-team | Repo: marimo}]
\textbf{Core Maintainer}: 
Project Maintainers lead the technical development of the marimo project, and they are the ultimate authority on the direction of the marimo project. [...] 
This includes triaging issues, proposing and reviewing pull requests, and updating continuous integration as needed. [...]
\end{tcolorbox}

In \texttt{vitessio/vitess}~\cite{vitess}, maintainers are entrusted with ongoing development and strategic alignment while remaining accountable through peer review and voting procedures. In \texttt{rook/rook}~\cite{rook}, maintainers combine hands-on development with community representation, issue triage, and participation in steering meetings. Likewise, \texttt{grafana/tempo}~\cite{tempo} defines maintainers as both technical leads and mediators of consensus.

\subsubsection{Committers and Reviewers: \textit{The Gatekeepers}}

Committers and reviewers show a narrower but essential set of skills. On the one side, committers are tightly focused on programming (T1), but also commonly involve management and interpersonal skills, such as coordinating merges, communicating changes, and aligning contributions with the project's goals. Their function is crucial, as they have direct access to the project's resources and can make changes without submitting patches. The \texttt{kopia/kopia} project~\cite{kopia} describes this commit-then-review process as efficient for trusted contributors, while their work continues to be reviewed by the community before acceptance.

\begin{tcolorbox}[title={Org: kopia | Repo: kopia}, breakable]
\textbf{Committer}: 
Committers are community members who have shown that they are committed to the project's continued development through ongoing engagement with the community. Committership allows contributors to more easily carry on with their project related activities by giving them direct access to the project’s resources. That is, they can make changes directly to project outputs, without having to submit changes via patches [...]
\end{tcolorbox}

Reviewers concentrate almost exclusively on code review (CT3) and analytical skills (P3). Their narrow scope underscores a quality-assurance orientation. The \texttt{distribution/distribution}~\cite{distribution} reviewers' votes are required to merge changes, making them guardians of project quality.

\begin{tcolorbox}[breakable, title={Org: distribution | Repo: distribution}]
\textbf{Reviewer}: 
A reviewer is a core role within the project. They share in reviewing issues and pull requests and their LGTM counts towards the required LGTM count to merge a code change into the project. [...]
\end{tcolorbox}

\subsubsection{Steering and Owners: \textit{The Strategists}}

In roles with greater strategic focus, steering and owner roles, technical skills are largely absent. Instead, they concentrate the management (M1, M2, M3), communication (IS2), and external relations (ER1, ER4). These roles embody governance as strategy rather than operation: they set direction, define policies, and coordinate with stakeholders. For example, the \texttt{crossplane/crossplane}~\cite{crossplane} Steering Committee role is tasked not with coding, but with \textit{"owning the overall charter and direction"} of the project, a responsibility reflected in the heavy management skills.

\begin{tcolorbox}[breakable, title={Org: crossplane | Repo: crossplane}]
\textbf{Steering Committee}: 
The Crossplane Steering Committee oversees the overall health of the project. Its made up of members that demonstrate a strong commitment to the project with views in the interest of the broader Crossplane community. Their responsibilities include: (i) own the overall charter, vision, and direction of the Crossplane project (ii) define and evolve project governance structures and policies, including project roles and how collaborators become maintainers. [...]
\end{tcolorbox}

Owners and founders, often associated with the benevolent dictator for life (BDFL) model~\cite{redhat_governance_2020}, exhibit skills distributions similar to steering roles, strong in management, communication, and external relations. Unlike steering committees, whose power is collective, founders centralize decision-making. For instance, in the \texttt{zellij-org/zellij} project~\cite{zellij}, the BDFL is responsible for strategic decisions on finance and collaborations, retaining the veto power as a safeguard for coordinating community decisions.

\begin{tcolorbox}[breakable, title={Org: zellij-org | Repo: zellij}]
\textbf{Benevolent Dictator for Life}: 
[...] is in charge of steering the project and making large decisions regarding finances and collaboration. He will bring such decisions to the group when possible in order to hear all dissenting opinions, but the ultimate decision is his. He also has veto power over decisions made by the group. Since we strive to make decisions by consensus, this power is to be used only as a last resort.
\end{tcolorbox}

\subsubsection{Triage: \textit{The Coordination and Production}}
The triage role shows a narrow skill footprint. The graphs indicate loading on bug triaging (CT1) and communication (IS2). This matches the function of triagers as translators: they take the noisy inflow of bug reports and transform them into structured, actionable work items. The \texttt{Prometheus-operator/prometheus-operator}~\cite{prometheusoperator} triage team captures this role precisely. Triagers are given GitHub permissions to adjust issues, allowing developers to focus on actual fixes.

\begin{tcolorbox}[breakable, title={Org: prometheus-operator | Repo: prometheus-operator}]
\textbf{Triage Team}: Contributors who have the Member role and Triage GitHub permission on the prometheus-operator organization and all projects within it, allowing them to modify GitHub issues and PRs statuses and labels. [...]
\end{tcolorbox}

\subsubsection{Specialized and Symbolic Roles: Community Advocate, Emeritus}

Beyond the central clusters, several roles carry symbolic weights. These positions show how projects carve out niches, delegate community-facing duties, or institutionalize memory.

\textbf{Community Advocate: \textit{The public voice}}. The community advocate role is one of the clearest cases of a non-technical skill profile. As shown in Figure \ref{fig:skillDistribution}, advocate skills are almost exclusively related to communication (IS2), external relations (ER2), and community management (M1). This suggests that advocates are expected to convey the project's identity through various channels, including blogs, social media, events, and partnerships. The \texttt{Rdatatable/data.table}~\cite{datatable} governance exemplifies this orientation. The advocate position is tasked with maintaining a blog, coordinating events, and ensuring visibility. 

\begin{tcolorbox}[breakable, title={Org: Rdatatable | Repo: data.table}]
\textbf{Community Advocate}: 
Community Engagement Coordinator. An individual who is involved in the project but does not also occupy the Committer or CRAN Maintainer role. In charge of maintaining The Raft blog, preparing Seal of Approval Applications, addressing Code of Conduct violations, and planning social or community events. [...]
\end{tcolorbox}

\textbf{Emeritus Maintainers: \textit{Memory and Symbolism}}. Unlike other roles, the Emeritus Maintainer category shows no measurable technical, managerial, or interpersonal skills. Emeritus roles institutionalize recognition, preserving the legacy of past contributors and ensuring continuity of identity even after active involvement has ceased. The \texttt{grafana/tempo}~\cite{tempo} project illustrates this. Here, emeritus status carries symbolic weight, signaling gratitude and respect for past leaders.

\begin{tcolorbox}[breakable, title={Org: grafana | Repo: tempo}]
\textbf{Emeritus Maintainer}: 
Emeritus maintainers are former maintainers who no longer work directly on Tempo on a regular basis. We respect their former contributions by giving them the Emeritus Maintainer title. This is honorary only and confers no responsibilities or rights regarding the Tempo project. [...]
\end{tcolorbox}

\subsubsection{Users: \textit{The Silent Majority}}
Finally, the "User" role appears almost deceptively simple in the graph, with light references to bug reporting (CT2) and communication (IS2). The \texttt{bpftrace/bpftrace}~\cite{bpftrace} project captures this ethos: users require no qualifications, but their presence justifies the entire endeavor. In governance terms, users are symbolic placeholders for the external audience that validates the project's value.

\begin{tcolorbox}[breakable, title={Org: Bpftrace | Repo: bpftrace}]
\textbf{User}: 
Users are community members who have a need for the project. They are the most important members of the community, and without them the project would have no purpose. Anyone can be a user; there are no special requirements. [...]
\end{tcolorbox}

\section{Discussion}

The preceding analysis detailed how OSS projects formalize participation through governance artifacts that define roles, responsibilities, and decision rights. In this section, we move from describing these patterns to interpreting what they reveal about organization and authority in open collaboration. 

\subsection{Governance as Textually Observable Infrastructure}

Our study demonstrates that governance in OSS projects is not only enacted through participation but also documented through written artifacts that make authority visible and comparable. The governance files analyzed here define who decides, who acts, and under what conditions, transforming what might otherwise remain tacit community norms into explicit, inspectable rules. 

By treating these artifacts as primary data, it is possible to analyze them through their textual form, much like source code or documentation. This perspective makes governance empirically observable and enables cross-project comparison. The method does not claim that written documents capture the entirety of practice, but rather that they provide a stable layer where organizational logic becomes legible.

\subsection{The Drift Between Codification and Practice}
While governance artifacts aim to clarify how participation is structured, our analysis reveals significant variation and ambiguity in how roles are described across projects. The same role title can denote very different responsibilities, while distinct titles often encode nearly identical duties. Some governance files omit key elements, such as promotion criteria or voting rules, leaving authority boundaries partially defined. 

\subsection{Governance as Layered System}
Our results show that governance in OSS projects tends to organize responsibilities into two broad layers. The organizational layer includes roles such as Owners, Chairs, and Steering Committees, which concentrate on strategic coordination, decision-making, and alignment with external stakeholders. The operational layer includes Contributors, Reviewers, and Triagers, who carry out the daily technical work. 

This division appears directly in governance documents, where responsibilities are grouped by scope rather than by seniority. However, the same documents also show an overlap and interdependence between layers. Maintainers, in particular, frequently act as the connective tissue between strategic and operational domains, embodying both coordination and production.

\subsection{The Maintainer Paradox}
Across projects, governance documents consistently frame maintainers as technical stewards while also assigning community-facing duties. This hybridization of authority produces what we call the \textit{Maintainer Paradox:} governance artifacts centralize power in those meant to distribute it. Our skill-mapping data reveals maintainers as the most multidimensional role, spanning technical, managerial, and interpersonal domains.

This observation reinforces concerns about bottlenecks and burnout in OSS leadership~\cite{guizani2021long, raman2020stress}. By coupling organizational and operational authority within a small group, projects may embed dependency, burnout and turnover into their governance.

\subsection{Implications for Governance Design}

Our analysis reveals how governance documents encode expectations, authority, and participation within OSS projects. From these observations, several implications emerge for those designing or revising governance models. First, projects should \textbf{make responsibilities explicit}. Governance files frequently describe similar functions under different titles, which can obscure accountability and confuse contributors. Clarifying what each role entails improves shared understanding within the community.

Second, projects need to be \textbf{aware of composite responsibilities}. Maintainers often accumulate multiple responsibilities, where the concentration of duties may hidden dependencies and potential burnout. Documenting these overlaps makes workload more visible to community. Third, the \textbf{inclusion of symbolic and communicative roles} underscores that governance extends beyond code production. These position reflect contributions that sustain community identity, cohesion, and continuity. Acknowledging these roles within governance structures reinforces the social dimension of sustainability.

Third, this work suggests that governance artifacts themselves constitute a \textbf{rich and underutilized source of empirical evidence}. By treating governance as textually observable infrastructure, researchers can systematically compare how authority and responsibility are formalized across projects. This perspective complements interview- or activity-based studies by providing a stable, inspectable layer where institutional logic is explicitly recorded.

Beyond practice, our findings contribute methodologically and theoretically to the study of governance in open collaboration. By introducing a role-centric analytical lens grounded in Institutional Grammar, we demonstrate how governance rules can be systematically decomposed into scopes, privileges, obligations, and transition criteria. This structured representation enables consistent comparison across heterogeneous projects, moving analysis beyond informal titles or anecdotal accounts. By applying this approach across repositories, licenses, and ecosystems, our study provides comparative evidence that supports theory-building about how OSS communities formalize authority, distribute responsibilities, and evolve their institutional structures over time.

\section{Study Design Trade-Offs and Limitations}
Following Robillard et al.~\cite{robillard2024communicating}, we report the key design decisions that shaped this study, the alternatives we considered, and their implications for interpretation. This approach highlights deliberate methodological reasoning and transparency instead of treating limitations as methodological flaws.

Our first major decision concerned the sampling frame and repository ranking. We stratified repositories by license type and ranked them by the number of GitHub stars within each category. Other sampling strategies (e.g., random or activity-based selection, contributor thresholds, or curated foundation lists) were considered but ultimately set aside. License stratification ensured broad ecosystem coverage, while star counts provided a reproducible, API-accessible indicator of visibility. 

However, as Borges and Valente~\cite{borges2018s} note, stars often reflect social signaling or appreciation rather than governance maturity. This means our dataset may over-represent socially prominent projects, and our findings should be interpreted as characterizing highly-visible communities that codify their governance structures, rather than all OSS projects.

A second decision concerned how we detected governance documents. We chose a narrow, reproducible heuristic, automatically identifying files whose names contained the term "governance" at any depth in the repository tree. Broader heuristics, or manual exploration of documentation sites and wikis, could have increased recall but by prioritizing a single, transparent rule, we ensured that detection remained verifiable and aligned with GitHub's "community health file" convention. The trade-off is that we likely excluded projects that govern informally or host their rules outside the repository.

We also made a key decision about the scope of textual evidence to analyze. We restricted our extraction to explicit statements found in governance files, marking elements such as privileges or responsibilities as absent when not directly expressed. Broader approaches (e.g., triangulating with issue discussions or interviews) could have captured tacit norms, but would have required substantial additional effort and time for data collection and interpretation. Our results, therefore, characterize governance as it is written, not necessarily as it is enacted.

In representing governance roles, we adopted a schema inspired by Institutional Grammar~\cite{crawford1995grammar, siddiki2011dissecting, sen2022cui}, decomposing each role into its scope, privileges, obligations, and promotion/demotion criteria. We considered looser thematic approaches and matrix-style responsibility models, but opted for the Institutional Grammar~\cite{crawford1995grammar, siddiki2011dissecting, sen2022cui} structure because it enables comparability across projects and aligns with rule-based representations of authority. While this schema standardizes analysis, it may fragment holistic responsibilities or understate the nuances of hybrid roles.

Finally, we initially applied unsupervised clustering to group roles, but found that composite or overlapping roles produced unstable results. We therefore pivoted to manual interpretive clustering, conducted collaboratively among authors and refined through consensus discussions. This increased interpretive judgment but allowed a richer understanding of how projects blend technical and organizational expertise. Our analysis emphasizes descriptive and comparative interpretation; we deliberately avoid linking governance forms to project outcomes such as retention or success, as they would require different data and design strategies.

\section{Conclusion}

This study examined how OSS projects define, structure, and differentiate roles through their written governance artifacts. By treating governance not as a static organizational chart but as a living textual system, we revealed how communities define authority, responsibility, and participation in writing. Governance files such as \texttt{GOVERNANCE.md} serve as socio-technical blueprints that document institutional arrangements that make participation pathways visible. 

Through a multi-project analysis, we identified recurrent patterns in how roles are articulated and responsibilities distributed. Projects consistently distinguish between organizational and operational tiers of work, but the distinction often overlaps in practice, especially through composite roles, such as the Maintainer. These hybrid roles embody what we termed the Maintainer Paradox. Other findings exposed symbolic and communicative functions that extend governance beyond code to encompass identity, legitimacy, and continuity.

Beyond the descriptive patterns, our results show that governance artifacts materialize the community's evolving balance between autonomy and structure. Written rules shape participation by defining what is legitimate to do and who is authorized to decide. In this sense, governance operates as a textual infrastructure that organizes collaboration without necessarily centralizing it.


\section{Replication Package}
To foster transparency, we provide a replication package with all the materials used in this study~\cite{paper_replication}. The package includes the complete dataset of governance files, role extractions, coding templates, and analytical scripts we used to generate the results presented in this paper. Instructions for reproducing the study are included in a \texttt{README.md} file. The dataset and code are available to encourage reuse and extension by the community. 

\begin{acks}
This work was supported by the National Science Foundation grant \#2303612. Also, CNPq grants \#314797/2023-8  and \#443934/2023-1

\end{acks}

\bibliographystyle{ACM-Reference-Format}
\bibliography{software}

\end{document}